\documentstyle[epsf]{lya2}
\begin{document}

\load{\scriptsize}{\sc}
\def\ion#1#2{\rm #1\,\sc #2}
\def\HI{{\ion{H}{i}}}
\def\HII{{\ion{H}{ii}}}
\def\GI{{\ion{He}{i}}}
\def\GII{{\ion{He}{ii}}}
\def\GIII{{\ion{He}{iii}}}
\def\MH{{{\rm H}_2}}
\def\Hp{{{\rm H}_2^+}}
\def\Hm{{{\rm H}^-}}

\def\dim#1{\mbox{\,#1}}

\def\figdir{/coma10/gnedin/SCR/FIGS/LYA2}
\def\figname#1{lya2.#1}

\title[Column density distribution]
{Probing the universe with the Lyman-alpha forest: \\II.
The column density distribution}

\author[Gnedin]{Nickolay Y.\ Gnedin
\\ Department of Astronomy, University of California, Berkeley, 
CA 94720
\\E-mail: \sl gnedin@astron.berkeley.edu;
http://astron.berkeley.edu/$\sim$gnedin
}

\maketitle

\begin{abstract}
I apply the well controlled Hydro-PM approximation of Gnedin \& Hui to
model the column density distribution of the Lyman-alpha forest for
25 different flat cosmological scenarios, including variants of the standard
CDM, tilted CDM, CDM with a cosmological constant, and CHDM models.
I show that within the accuracy of the HPM approximation the slope
of the column density distribution reflects the degree of nonlinearity
of the cosmic gas distribution and is a function of the rms 
linear density fluctuation at the characteristic filtering scale only. 
The amplitude
of the column density distribution, expressed as the value for the ionizing 
intensity, is derived as a function of the cosmological parameters 
(to about 40\%
accuracy). The observational data are currently consistent with the value 
for the ionizing intensity being constant in the redshift interval
$z\sim2-4$.
\end{abstract}


\section{Introduction}

Recent years witnessed a rapid development in our understanding of the
Lyman-alpha forest, a phenomenon, potentially offering cosmologists the
most accurate probe of the Intergalactic Medium (IGM) at intermediate
cosmological redshifts ($z\sim2-4$). On one hand, observations with the
Keck telescopes 
(see, for example, Hu et al.\ 1995; Lu et al.\ 1996;
Cristiani et al.\ 1996; Kirkman \& Tytler 1997;
Kim et al.\ 1997; and D'Odorico et al.\ 1997) provided an impressive amount 
of high quality data. On the other hand, cosmological hydrodynamic simulations 
\cite{CMOR94,ZAN95,HKWM96,MCOR96,WB96,ZANM97}
offered a theory
whose {\it a priori\/} predictions can now not only be quantitatively
compared to the observational data, but also, and perhaps even more
importantly, be used to constraint the cosmology with the Lyman-alpha
forest observations.

While a possibility still remains that the true nature of the Lyman-alpha 
is still
uncovered, or that there exist a genuine population of individual absorbers
\cite{BS65,A72,B81,OI83,IO86,R86,I86,R88,BSS88},
this paper is solely based on the
theory that Lyman-alpha absorption arises from the fluctuating IGM, and the
fluctuations in the IGM are small-scale siblings of larger-scale
structures, that one sees in the distribution of galaxies in the sky,
as argued for by cosmological hydrodynamic simulations.

However, cosmological hydrodynamic simulations may be the worst
theoretical tool 
one can imagine. They require enormous (compared to the vast
majority of theoretical methods) computational resources, and at the same
time demand high expertise from a researcher, who ought to understand
a vast number of subtle numerical and modeling biases arising from
using such a complicated tool as a cosmological hydrodynamic code.
The development of the code takes months, if not years, of human labor, and
currently there exist only a few groups who engage in modeling the
Lyman-alpha forest with hydrodynamic simulations. This situation will
change in the future, when supercomputers are faster and more
widely available, and when standard hydrodynamic codes become
publicly accessible, but the current situation vows for the use of
approximate methods, which sacrifice the accuracy for the sake of
simplicity and speed. The simplicity and speed are especially important
because currently there exist a large variety of cosmological models that
need to be tested against the observational data; while one or two specific 
cosmological models can be best studied with the hydrodynamic simulations,
it is currently beyond reason to investigate dozens of models with the
hydrodynamic simulations.

Up to date, two different approximations have been used for the purpose of
modeling the Lyman-alpha forest: the lognormal 
approximation \cite{BBC92,GH96,BD97} and the 
Zel'dovich approximation \cite{DS77,MG90,HGZ97}. While both approximations
are very efficient and much easier to implement than a cosmological
hydrodynamic simulation, they are {\it uncontrolled\/} approximations, i.e.\
the range of their applicability and their accuracy is not well established.
As a result, their predictions cannot be directly used in a quantitative
comparison with the observational data, since it is not clear what
the error is to assign to the theoretical prediction based on an
approximation (since an approximation is used, it has to have an error
associated with it).

In this paper I use a different approximation, called HPM, and developed
in Gnedin \& Hui \shortcite{GH97}.
In its essence, the HPM method uses a simple Particle-Mesh
(PM) solver modified to account for the effect of gas pressure. 
Because there exist a tight correlation between gas temperature/pressure 
and gas density (``equation of state'') in the low density IGM in the
mildly nonlinear regime (cosmic overdensity $\delta\la10$), it is
possible to compute the gas temperature and pressure at every point
directly from the value of the cosmic gas density at this point. Thus,
there is no need to introduce a special equation for the gas temperature
as in a full hydrodynamic solver. As a result, the HPM approximation
is only about 25\% slower (due to the overheard of computing the equation of
state) than a simple PM solver, and substantially faster than a full
hydrodynamic solver (due to both fewer computations at each time-step
and fewer time-steps), while delivering results which are accurate to 
about 15\% in the point-by-point comparison with a full hydrodynamic
simulation of precisely the same cosmological model (for $\delta\la10$). 

Standard (i.e.\ unmodified to account for the gas pressure) PM simulations 
have been also used to address properties of the Lyman-alpha forest
\cite{PMK95,MPKR96,W97}. While PM simulations are most likely more accurate
than lognormal or Zel'dovich approximations, they still cannot quite reach
the level of accuracy achieved by the HPM method. For example, as has been
shown by Gnedin \& Hui \shortcite{GH97}, a standard PM simulation
predicts about a factor of 3 lower densities than the full hydrodynamic
simulation; it is possible to improve upon the PM simulation by appropriate
smoothing of the initial conditions, but even in this case a PM simulation is
only able to reproduce the gas density to about 30-40\% for $\delta\la10$, 
whereas HPM reaches a 15\% accuracy.

In general, it does not seem to be much rationale in using the standard
PM simulation instead of HPM, because 25\% increase in the CPU time in the  
HPM approach compared to simple PM buys at least a factor of 2 in the error 
reduction. Also,
from a technical point of view, it does not require much human effort
to convert a PM code into an HPM code.

Having chosen the tool to study the Lyman-alpha forest, it is most natural
to start with investigating the column density distribution, since it
is the quantity best known observationally (to within about 10-15\%).
Since the main advantage of using the HPM approximation is its speed,
I concentrate in this paper on analysing the properties of the column
density distribution of the Lyman-alpha forest for a large set of
different cosmological models (25 altogether) with the main goal to
understand what physical parameters the column density distribution
depends upon. Testing of specific cosmological models is only
a secondary goal of this paper.

I begin my analysis with discussing the models and characteristic scales
in Section 2. In the same section I also discuss another approximation I have
to adopt to conclude my study, and carefully investigate the total error
induced by using approximations rather than an exact method (in this case a
cosmological hydrodynamic simulation). The results are presented in Section 3,
and Section 4 concludes the paper with the discussion.

\section{Method}

\subsection{Cosmological models}

\def\hidea#1{}
\def\hideb#1{#1}
\begin{table*}
\begin{minipage}{\hsize}
\caption{Cosmological Models}
\label{tabmod}
\begin{tabular}{lllllllllll\hideb{ll}llc}
\noalign{\hrule\smallskip}
Model & $\Omega_0$ & $\Omega_b$ & $\Omega_\Lambda$ & $\Omega_{\nu}$ & $h$ & $n$ & $\sigma_8$ & 
$\sigma_{34}$ & $T_{4,\rm\, rei}$ & $z_{\rm rei}$ &\hideb{ $\gamma$ & $T_4$ &} $\sigma_F$ & $\sigma_{F,{\rm BOX}}$ & COBE \\
\noalign{\smallskip\hrule\smallskip}
SCDM.1A & 1    & 0.05  & 0    & 0    & 0.5  & 1    &  0.8  & 2.22 & 2.5 &  7 & \hidea{1.407 1.16 2.10 1.93  1.441 1.06 2.20 2.00}\hideb{  1.49 & 0.91 &} 2.39 & 2.15 \hidea{1.526 0.76 2.67 2.38} &  \\
{\bf SCDM.2A} & 1    & 0.05  & 0    & 0    & 0.5  & 1    &  0.7  & 1.95 & 2.5 &  7 & \hidea{1.407 1.16 1.84 1.68  1.441 1.06 1.93 1.75}\hideb{  1.49 & 0.91 &} 2.09 & 1.88/2.00 \hidea{1.526 0.76 2.33 2.08} &  \\
{\bf SCDM.2D}$^{\rm a}$
        & 1    & 0.05  & 0    & 0    & 0.5  & 1    &  0.7  & 1.95 & 2.5 &  7 & \hidea{1.407 1.16 1.84 1.68  1.441 1.06 1.93 1.75}\hideb{  1.49 & 0.91 &} 2.09 & 1.88/2.00 \hidea{1.526 0.76 2.33 2.08} &  \\
SCDM.2E$^{\rm a}$
        & 1    & 0.05  & 0    & 0    & 0.5  & 1    &  0.7  & 1.95 & 2.5 &  7 & \hidea{1.407 1.16 1.84 1.68  1.441 1.06 1.93 1.75}\hideb{  1.49 & 0.91 &} 2.09 & 1.88 \hidea{1.526 0.76 2.33 2.08} &  \\
SCDM.2G & 1    & 0.05  & 0    & 0    & 0.5  & 1    &  0.7  & 1.95 & 1.5 &  7 & \hidea{1.436 0.77 1.96 1.82  1.468 0.70 2.06 1.89}\hideb{  1.51 & 0.61 &} 2.23 & 2.04 \hidea{1.544 0.53 2.49 2.26} &  \\ 
SCDM.2H & 1    & 0.05  & 0    & 0    & 0.5  & 1    &  0.7  & 1.95 & 4.0 &  7 & \hidea{1.382 1.74 1.73 1.56  1.417 1.56 1.81 1.62}\hideb{  1.46 & 1.32 &} 1.96 & 1.74 \hidea{1.509 1.09 2.19 1.92} &  \\ 
\noalign{\smallskip}
SCDM.2J & 1    & 0.05  & 0    & 0    & 0.5  & 1    &  0.7  & 1.95 & 2.5 & 15 & \hidea{1.599 0.88 1.62 1.45  1.602 0.83 1.73 1.54}\hideb{  1.61 & 0.74 &} 1.92 & 1.70 \hidea{1.609 0.65 2.19 1.92} &  \\
SCDM.2L & 1    & 0.05  & 0    & 0    & 0.5  & 1    &  0.7  & 1.95 & 2.5 &  5 & \hidea{1.203 1.66 2.13 2.00  1.255 1.47 2.16 2.01}\hideb{  1.33 & 1.21 &} 2.26 & 2.00 \hidea{1.403 0.97 2.46 2.22} &  \\
SCDM.3A & 1    & 0.05  & 0    & 0    & 0.5  & 1    &  0.6  & 1.67 & 2.5 &  7 & \hidea{1.407 1.16 1.58 1.44  1.441 1.06 1.65 1.50}\hideb{  1.49 & 0.91 &} 1.79 & 1.62 \hidea{1.526 0.76 2.00 1.79} &  \\
{\bf SCDM.4A} & 1    & 0.05  & 0    & 0    & 0.5  & 1    &  0.5  & 1.39 & 2.5 &  7 & \hidea{1.407 1.16 1.31 1.20  1.441 1.06 1.38 1.25}\hideb{  1.49 & 0.91 &} 1.49 & 1.35/1.43 \hidea{1.526 0.76 1.67 1.49} &  \\
\noalign{\smallskip}
{\bf SCDM.5A} & 1    & 0.05  & 0    & 0    & 0.5  & 1    &  1.2  & 3.34 & 2.5 &  7 & \hidea{1.407 1.16 3.16 2.89  1.441 1.06 3.30 3.01}\hideb{  1.49 & 0.91 &} 3.58 & 3.23/3.52 \hidea{1.526 0.76 4.00 3.57} & $\surd$ \\
TCDM.2A & 1    & 0.05  & 0    & 0    & 0.5  & 0.9  &  0.83 & 1.95 & 2.5 &  7 & \hidea{1.407 1.16 1.82 1.63  1.441 1.06 1.91 1.84}\hideb{  1.49 & 0.91 &} 2.09 & 1.84 \hidea{1.526 0.76 2.34 2.04} & $\surd$ \\
TCDM.4A & 1    & 0.05  & 0    & 0    & 0.5  & 0.7  &  1.16 & 1.95 & 2.5 &  7 & \hidea{1.407 1.16 1.80 1.53  1.441 1.06 1.91 1.60}\hideb{  1.49 & 0.91 &} 2.10 & 1.74 \hidea{1.526 0.76 2.38 1.95} &  \\
LCDM.1A & 0.35 & 0.05  & 0.65 & 0    & 0.7  & 1    &  0.94 & 1.85 & 2.5 &  7 & \hidea{1.476 1.50 1.72 1.54  1.504 1.40 1.79 1.59}\hideb{  1.53 & 1.25 &} 1.94 & 1.69 \hidea{1.564 1.09 2.15 1.85} & $\surd$ \\
LCDM.1C & 0.35 & 0.05  & 0.65 & 0    & 0.7  & 1    &  0.94 & 1.85 & 1.5 &  5 & \hidea{1.295 1.16 2.14 2.00  1.356 1.08 2.16 2.00}\hideb{  1.43 & 0.96 &} 2.26 & 2.06 \hidea{1.493 0.83 2.44 2.19} & $\surd$ \\
\noalign{\smallskip}
LCDM.2A & 0.35 & 0.05  & 0.65 & 0    & 0.7  & 0.96 &  0.80 & 1.49 & 1.5 &  5 & \hidea{1.295 1.16 1.68 1.56  1.356 1.08 1.71 1.57}\hideb{  1.43 & 0.96 &} 1.80 & 1.62 \hidea{1.493 0.83 1.95 1.73} & $\surd$ \\
LCDM.3A & 0.35 & 0.03  & 0.65 & 0    & 0.7  & 1    &  1.04 & 2.14 & 2.5 &  7 & \hidea{1.435 1.28 2.01 1.81  1.467 1.17 2.10 1.88}\hideb{  1.51 & 1.02 &} 2.27 & 2.01 \hidea{1.543 0.87 2.53 2.20} & $\surd$ \\
LCDM.4A & 0.35 & 0.03  & 0.65 & 0    & 0.7  & 0.95 &  0.85 & 1.62 & 2.5 &  7 & \hidea{1.435 1.28 1.51 1.35  1.467 1.17 1.59 1.40}\hideb{  1.51 & 1.02 &} 1.72 & 1.50 \hidea{1.543 0.87 1.92 1.65} & $\surd$ \\
LCDM.5A & 0.4  & 0.036 & 0.6  & 0    & 0.65 & 1    &  1.02 & 2.12 & 2.5 &  7 & \hidea{1.439 1.29 1.99 1.80  1.470 1.18 2.08 1.87}\hideb{  1.51 & 1.04 &} 2.25 & 2.00 \hidea{1.546 0.89 2.50 2.20} & $\surd$ \\
CHDM.1A & 1    & 0.05  & 0    & 0.1  & 0.5  & 1    &  0.91 & 2.13 & 2.5 &  7 & \hidea{1.407 1.16 1.99 1.80  1.441 1.06 2.08 1.88}\hideb{  1.49 & 0.91 &} 2.26 & 2.02 \hidea{1.526 0.76 2.53 2.23} & $\surd$ \\
\noalign{\smallskip}
CHDM.2A & 1    & 0.05  & 0    & 0.15 & 0.5  & 1    &  0.85 & 1.69 & 2.5 &  7 & \hidea{1.407 1.16 1.56 1.40  1.441 1.06 1.64 1.46}\hideb{  1.49 & 0.91 &} 1.78 & 1.57 \hidea{1.526 0.76 1.99 1.73} & $\surd$ \\
CHDM.3A & 1    & 0.05  & 0    & 0.2  & 0.5  & 1    &  0.81 & 1.35 & 2.5 &  7 & \hidea{1.407 1.16 1.23 1.08  1.441 1.06 1.30 1.13}\hideb{  1.49 & 0.91 &} 1.41 & 1.22 \hidea{1.526 0.76 1.58 1.35} & $\surd$ \\
CHDM.4A & 1    & 0.05  & 0    & 0.1  & 0.65 & 1    &  1.21 & 3.29 & 2.5 &  7 & \hidea{1.428 1.24 3.06 2.80  1.461 1.14 3.20 2.91}\hideb{  1.50 & 0.99 &} 3.46 & 3.12 \hidea{1.540 0.84 3.86 3.44} & $\surd$ \\
CHDM.5A & 1    & 0.05  & 0    & 0.2  & 0.65 & 1    &  1.11 & 2.10 & 2.5 &  7 & \hidea{1.428 1.24 1.91 1.69  1.461 1.14 2.01 1.76}\hideb{  1.50 & 0.99 &} 2.19 & 1.89 \hidea{1.540 0.84 2.44 2.09} & $\surd$ \\
CHDM.6A & 1    & 0.07  & 0    & 0.1  & 0.6  & 0.9  &  0.72 & 1.55 & 2.5 &  7 & \hidea{1.449 1.34 1.41 1.26  1.480 1.24 1.48 1.31}\hideb{  1.52 & 1.09 &} 1.61 & 1.41 \hidea{1.551 0.94 1.81 1.56} & $\surd$ \\
\noalign{\smallskip\hrule\medskip}
\end{tabular}
\end{minipage}
\smallskip
$^{\rm a}$ A different random realization of the SCDM.2A model.\hfill\break
\end{table*}
With the help of HPM, it is possible to simulate a large set of cosmological 
models with only modest computational resources. Table \ref{tabmod} lists
all 25 simulations considered in this paper. Simulations have been performed
with $256^3$ particles on a $256^3$ mesh. An average $256^3$
simulation takes about 40
CPU-hours on a R10K processor. In addition, 
4 simulations have been rerun with $512^3$ particles on a $512^3$ mesh,
to verify numerical convergence. Those four models are shown in bold
face in Table \ref{tabmod}. Each of $512^3$ simulations requires about
450 CPU-hours on a R10K processor, and it is impractical at the moment
to perform all simulations listed in Table \ref{tabmod} with the
$512^3$ mesh. I therefore use the $512^3$ simulations to investigate
numerical convergence and accuracy of the results obtained with $256^3$
models.

Placement of a simulation
box has been chosen so as to insure that the characteristic scale over which
linear baryon fluctuations are smoothed, the so called {\it filtering
scale\/} \cite{GH97}, is resolved with at least 3 cells. This dictates
the comoving cell size of $10h^{-1}\dim{kpc}$ for $\Omega_0=1$ models
and $15h^{-1}\dim{kpc}$ for LCDM models, with the exception of the
SCDM.2K and SCDM.2L models, whose low reionization temperature and
late reionization (and therefore a small filtering
scale) required to use the cell size of $7h^{-1}\dim{kpc}$.

Cosmological parameters and labels for all models are summarized in Table 
\ref{tabmod}. Here $\Omega_0$ is the total matter density parameter,
$\Omega_b$ is the baryon density parameter, $\Omega_\Lambda$ is the 
cosmological constant density parameter, and $\Omega_{\nu}$ is the density
parameter in massive neutrinos. As usual, $h$ denotes the Hubble constant
in units of $100\dim{km}/\dim{s}/\dim{Mpc}$, $n$ is the power-law index
of the primordial spectrum of density fluctuations ($n=1$ is the scale-free
Harrison-Zel'dovich spectrum), and $\sigma_8$ is the rms 
top-hat density fluctuation on
$8h^{-1}\dim{Mpc}$ at $z=0$,
\[
	\sigma_8^2 \equiv \int_0^\infty {k^2 dk\over 2\pi^2} 
	P(k) W_{\rm TH}^2(k\times8h^{-1}\dim{Mpc}).
\]

In order to simulate the evolution of the IGM under the HPM approximation,
an equation of state (i.e.\ the temperature-density relation) as a function 
of time ought to be specified. In general, the equation of state depends on 
the time evolution and spectral shape of the ionizing background, which is
a two-dimensional function. In order to limit the possible choice of
time-dependent equations of state to a manageable set, I adopt the power-law
equation of state in the form derived in Hui \& Gnedin \shortcite{HG97},
\begin{equation}
	T(\rho) = T_0(1+\delta)^{\gamma-1},
	\label{eos}
\end{equation}
where $\delta$ is the cosmic overdensity, and $T_0$ and $\gamma$ are
functions of redshift. Specifically, I adopt the evolution of $T_0$ and
$\gamma$ from Hui \& Gnedin \shortcite{HG97}, equations (17) and (19).
In this case the evolution of the equation of state is determined by only
two parameters, the epoch of sudden reionization $z_{\rm rei}$, and
the temperature of the gas just after reionization, $T_{\rm rei}$.
Those two parameters are shown in Table \ref{tabmod} for all models,
with the reionization temperature expressed in units of
$10^4\dim{K}$,
\[
	T_{4,\rm\, rei} \equiv T_{\rm rei}/10^4\dim{K},
\]
with reasonable values ranging from $T_{4,\rm\, rei}\sim1$ to 
$T_{4,\rm\, rei}\sim4$. For the purpose of this paper it is only
important that equation (\ref{eos}) gives {\it an\/} equation of state,
and by changing parameters $T_{\rm rei}$ and $z_{\rm rei}$ I can span
a reasonable range in possible equations of state. As will be shown
below, the specific form of the equation of state plays a relatively minor
role in determining the column density distribution of the Lyman-alpha
forest.

Given the specific form of the evolution of the equation of state, it is
then possible to compute the filtering scale as a function of redshift
for a given cosmological model. In general, the wavenumber $k_F$
corresponding to the filtering scale is given
by the following expression \cite{GH97}:
\begin{eqnarray}
        {1\over k_F^2(t)} & = & {1\over D_+(t)} \int_0^t dt^\prime 
        a^2(t^\prime)
        {\ddot{D}_+(t^\prime)+2H(t^\prime)\dot{D}_+(t^\prime) 
        \over k_J^2(t^\prime)} \nonumber \\
        & & \int_{t^\prime}^t{dt^{\prime\prime}\over a^2(t^{\prime\prime})},
        \label{kfaskj}
\end{eqnarray}
where $D_+(t)$ is the growing mode of linear fluctuations, and 
\begin{equation}
        k_J \equiv {a \over c_S}\sqrt{4\pi G\bar\varrho}
        \label{defkj}
\end{equation}
is the wavenumber corresponding to the Jeans scale. 
Here $c_S$ is the sound speed of the cosmic gas,
and $\bar\varrho$ is the total cosmic mean density.
For the fully ionized cosmic gas with the mean molecular weight
$\mu=0.6$ and with the equation of state given by (\ref{eos}),
$k_J$ can be expressed as
\begin{equation}
	k_J(z) = 8.44 \Omega^{1/2} h \dim{Mpc}^{-1}
	\left(1.5(1+z)\over\gamma T_4\right)^{1/2},
	\label{kjfor}
\end{equation}
where $T_4\equiv T_0/10^4\dim{K}$.
In general, the filtering scale is a complicated function of time,
but for a large set of reasonable equations of state it is about a 
half of the Jeans scale at $z\sim3$, $k_F\sim 2k_J$. 
It is therefore useful to introduce a measure of nonlinearity of the
gas distribution of a given cosmological model at $z=3$,
$\sigma_{34}$, defined as
\begin{equation}
	\sigma_{34}^2 \equiv \int_0^\infty {k^2 dk\over 2\pi^2} 
	P(k,z=3) \exp(-2k^2/k_{34}^2),
	\label{sig34def}
\end{equation}
where
\begin{equation}
	k_{34} \equiv 34 \Omega_0^{1/2}h \dim{Mpc}^{-1}
	\label{ktfdef}
\end{equation}
is the wavenumber corresponding to one half of the Jeans scale
for the cosmic gas with $T_4=1$ and $\gamma=1.5$ at $z=3$. The
particular choice of filtering was discussed in Gnedin \& Hui
\shortcite{GH97}. The
quantity $\sigma_{34}$ should therefore indicate an approximate degree 
of nonlinearity of the gas distribution in a given cosmological model
at $z\approx3$. A more accurate indicator is of course the rms linear density
fluctuation at the filtering scale,
\begin{equation}
	\sigma_{F}^2(z) \equiv \int_0^\infty {k^2 dk\over 2\pi^2} 
	P(k,z) \exp\left[-2k^2/k_{F}^2(z)\right],
	\label{sigfdef}
\end{equation}
and $k_F(z)$ is a function of time. While $\sigma_F$ is a more accurate
measure, it is also more difficult to compute (since it requires evaluation
of $k_F$). On the other hand, $\sigma_{34}$ is less accurate but 
easier to compute, and it can be useful
for quick, order-of-magnitude estimates.

Since a simulation is performed on a finite box, only a finite
range of scales is present in the simulation. Thus, the rms density
fluctuation computed in a simulation will be smaller than the one
obtained using the full range of scales. To be able to account for
this effect, I introduce another quantity, $\sigma_{F,{\rm BOX}}$,
similar to equation (\ref{sigfdef}), but with integration performed
over only a finite range of scales from $k_{\rm min}$ to $k_{\rm max}$
which are present in the simulation box:
\begin{equation}
	\sigma_{F,{\rm BOX}}^2(z) \equiv \int_{k_{\rm min}}^{k_{\rm max}}
	{k^2 dk\over 2\pi^2} 
	P(k,z) \exp\left[-2k^2/k_{F}^2(z)\right],
	\label{sigfboxdef}
\end{equation}
The small scale cut-off,
$k_{\rm max}$, is not that important because the exponential
filtering insures that, as long as the filtering scale $1/k_F$ is
resolved with at least three cell sizes ($k_F<1/(3l)$ where
$l$ is the cell size), any contribution to the integral at
$k\sim k_{\rm max}\equiv \pi/l$ (the Nyquist frequency)
is suppressed by a factor $\exp[-2(3\pi)^2]$, which is equal zero for any
practical purpose.

Values of $\sigma_{34}$,
$\sigma_F(z=2.85)$, and $\sigma_{F,{\rm BOX}}(z=2.85)$ 
are shown in Table \ref{tabmod} (for four models for which
both $256^3$ and $512^3$ simulations have been run, both values
of $\sigma_{F,{\rm BOX}}(z=2.85)$ are shown).
\hideb{Also, for reference purposes, I show values for $T_4$ and
$\gamma$ which define the equation of state of the IGM at $z=2.85$.}
The redshift value of $z=2.85$ corresponds to the epoch where most
abundant observation data on the column density distribution exist
\cite{HKCS95,KT97,KHCS97}.
The check mark in the last column of Table \ref{tabmod}
indicates whether a particular model is {\it COBE\/} normalized. The
{\it COBE\/} normalization was done using Bunn \& White \shortcite{BW97}
fits and assuming the gravity wave contribution for $n<1$ models.

In order to specify initial conditions for a simulation, a linear matter
transfer function needs to be computed. For my LCDM simulations I used
transfer functions computed using the linear gravity code
(existing in the literature fits are not sufficiently accurate due to
the higher baryonic fraction in LCDM models compared to 
$\Omega_0=1$ models); Hu
\& Sugiyama \shortcite{HS96} approximation was used to compute transfer
functions for SCDM and TCDM simulations, and Ma \shortcite{M96} fits
were used for CHDM simulations.

Finally, a few words about the choice of models. Apparently, only
{\it COBE\/} normalized models can be considered viable. The set of
SCDM models, which are not {\it COBE\/} normalized, is therefore used
as a kind of ``training set'', with which dependence on different parameters
is analysed in order to develop a general understanding of how the column
density distribution of the Lyman-alpha forest depends on the underlying
cosmology.

\subsection{The column density distribution}

The purpose of this paper is to investigate the column density distribution
of the Lyman-alpha forest, and I therefore need a way to compute the column
density distribution from the output of a simulation.
The standard way of accomplishing this is to generate synthetic spectra
and then fit Voigt profiles \cite{DHWK97}. However, the Voigt profile fitting
is extremely CPU intensive, and full Voigt profile analysis of results
of one simulation would require more computer time than the simulation
itself consumed. It is therefore feasible to use an approximate technique
and then properly account for the error induced.

Fortunately, there is such an approximate method called the ``Density-Peak
Ansatz'' (hereafter DPA), introduced in \cite{GH96,HGZ97}. The DPA 
approximation is based on the assumption that a single Lyman-alpha line
arises from a peak in the cosmic gas density along the line of sight 
to a distant quasar. Then,
given the value of the cosmic density $\rho\equiv1+\delta$ at the peak,
and the second derivative of the density along the line of sight
(the first derivative vanishes at the peak), a value for the column
density can be associated with the peak as follows:
\[
	N_{\HI} \equiv 1.45\times10^{13}\dim{cm}^{-2}T_4^{-0.7}
	\left(\Omega_bh^2\over0.0125\right)^2\left(0.5\over J_{21}
	\right)\left(1+z\over4\right)^5
\]
\begin{equation}
	\ \ 
	(1+\delta)^{2-0.7(\gamma-1)}
	\left[-(1-0.35(\gamma-1)){d^2\dim{ln}(1+\delta)\over dx^2}
	\right]^{-1/2},
	\label{ncdd}
\end{equation}
and $x$ being measured in Mpc. Here $J_{21}$ is the ionizing intensity,
\[
        J_{21} \equiv 
        {\int J_\nu \sigma_\nu d\nu/\nu\over \int \sigma_\nu d\nu/\nu}
        \times
        {1\over10^{-21}\dim{erg}\dim{cm}^{-2}\dim{s}^{-1}\dim{Hz}^{-1}
        \dim{sr}^{-1}},
        \label{J21def}
\]
where $J_\nu$ is the radiation intensity and $\sigma_\nu$ is the hydrogen
photoionization cross-section.

\def\capSP{
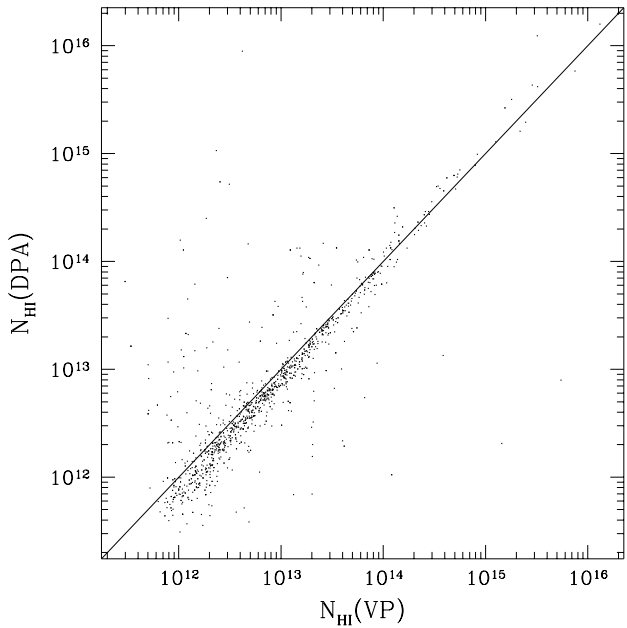
Scatter plot of column densities computed using the DPA approximation
vs column densities
obtained by Voigt profile fitting for the SCDM.2A model with $J_{21}=0.17$.
}
\begin{figure}
\par\centerline{%
\epsfxsize=1.0\columnwidth\epsfbox{\figname{figSP.ps}}}%
\caption{\label{figSP}\capSP}
\end{figure}
In order to test the DPA approximation, I have computed synthetic spectra
for random lines of sight through the simulation box for several models
at $z=2.85$ and fitted Voigt profiles to Lyman-alpha forest lines using
AUTOVP automatic Voigt profile fitting code \cite{DHWK97}, kindly provided
to me by Romeel Dave. In order to do that, I need to assume a value for
the ionizing intensity $J_{21}$, and also specify the signal-to-noise
ratio $S/N$ for Voigt profile fitting (the DPA column densities are 
independent of the adopted signal-to-noise ratio).
Figure \ref{figSP} shows the scatter plot of column 
densities from the DPA approximation versus column densities computed
using Voigt profile fitting for the SCDM.2A model assuming $J_{21}=0.17$
(which gives a reasonable fit to the observational data 
from Hu et al.\ 1995) and $S/N=50$. Note, that while there are absorption lines
on which the DPA approximation fails severely, in large the agreement is
remarkable, keeping in mind the simplicity of the DPA method. 

\def\capVP{
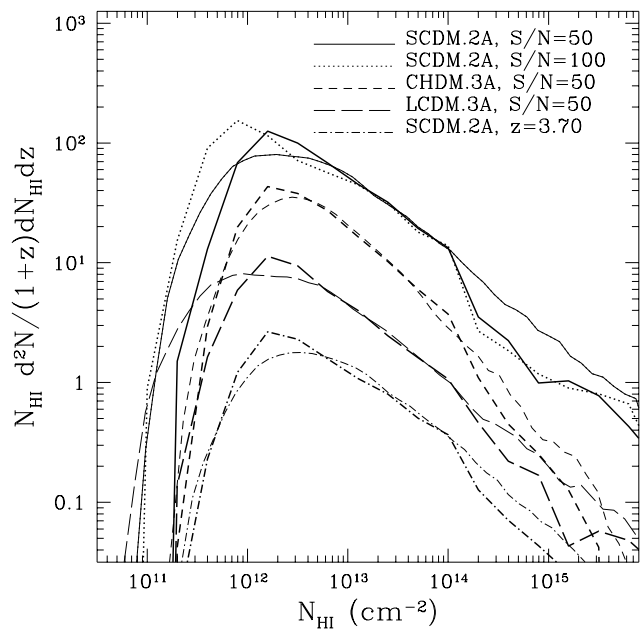
Column density distributions computed using the DPA method
(thin lines) and Voigt profile fitting (bold lines) for
the SCDM.2A model with $J_{21}=0.17$ and $S/N=50$ (solid lines),
the SCDM.2A model with $J_{21}=0.17$ and $S/N=100$ (dotted lines),
the CHDM.3A model with $J_{21}=0.25$ and $S/N=50$ (short-dashed lines),
the LCDM.3A model with $J_{21}=0.3$ and $S/N=50$ (long-dashed lines),
and
the SCDM.2A model with $J_{21}=0.25$ and $S/N=50$ at $z=3.70$
(dot-dashed lines). The last three models are plotted with the 
successive vertical offsets of 0.5 dex for clarity.
}
\begin{figure}
\par\centerline{%
\epsfxsize=1.0\columnwidth\epsfbox{\figname{figVP.ps}}}%
\caption{\label{figVP}\capVP}
\end{figure}
Column density distributions from the DPA method and Voigt profile fitting
for several models are shown in Figure \ref{figVP} with thin and bold lines
respectively. The solid line shows the SCDM.2A model with $J_{21}=0.17$
and $S/N=50$, while the dotted line shows the same model but with
$S/N=100$. I also show the CHDM.3A and LCDM.3A models on the same
plot, as well as the SCDM.2A model at $z=3.70$. 

One can see that DPA agrees very well (to about 3\%) with the column
density distribution from Voigt profile fitting for column densities
between about $10^{13}\dim{cm}^{-2}$ and $10^{14}\dim{cm}^{-2}$
independently of the signal-to-noise ratio in the Voigt profile fitting
procedure,
but fails at lower and higher column densities. At the lower end, the DPA
column density distribution falls off due to the finite resolution of 
the simulations; in this regime the DPA approximation becomes unreliable,
and, at the same time, the Voigt profile fitting technique becomes sensitive
to the assumed signal-to-noise ratio in the synthetic spectrum. 
One can expect, however,
that higher resolution (larger mesh size) simulations would improve 
agreement between the DPA and Voigt profile fitting for lower column 
densities.

For column densities
in excess of $10^{14}\dim{cm}^{-2}$, i.e.\ for {\it saturated lines\/},
the column density distribution computed using Voigt profile fitting shows
a feature, while the DPA column density distribution continues as a smooth
function of the column density. It is therefore plausible to presume that
the feature in the column density distribution obtained by Voigt profile 
fitting is
an artifact of the procedure rather than a change in intrinsic properties
of absorbers. However, since all available observational data were analysed
using the Voigt profile fitting
technique, the same artifact is present in data as well.
Thus, it seems reasonable to restrict the range of applicability of the DPA
method to column densities between about $10^{12.6}\dim{cm}^{-2}$ and 
$10^{14.0}\dim{cm}^{-2}$, where the agreement is excellent. As a side note
I point out here that this range of column densities is also within the
range of applicability of the HPM approximation, which gives accurate
results only for overdensities $\delta\la10$.

\subsection{The error budget}

The HPM approximation for the hydrodynamics of the low density IGM,
combined with the DPA approach, is a powerful and accurate method of
computing the column density distribution of unsaturated Lyman-alpha
absorption lines. Both approximation, the HPM, and the DPA, are
{\it controlled\/} approximations, i.e.\ their range of applicability
and their accuracy is well known. It is therefore possible to compute
the accuracy with which the column density distribution can be computed
using the HPM-DPA combination. 

\begin{table}
\caption{Error budget}
\label{taberr}
\medskip
\begin{tabular}{lcc}
\noalign{\hrule\smallskip}
Source of error & Slope & Amplitude  \\
\noalign{\smallskip\hrule\smallskip}
HPM is not exact &  3\% &  3\% \\
DPA is not exact &  3\% &  2\% \\
Finite box size  &  13(3)\% & 18(5)\% \\
\noalign{\smallskip\hrule\smallskip}
Total error      &  14(5)\% & 19(6)\% \\
\noalign{\smallskip\hrule}
\end{tabular}
\end{table}
It is now customary to approximate the column density distribution of the 
Lyman-alpha forest with the power law over a finite range of column densities.
Thus, I can convert the error in the computed column density distribution to
errors in the amplitude and the slope of the power law. There are three
different sources of error in the HPM-DPA calculation, which are summarized in 
Table \ref{taberr}. First, the HPM itself is not exact.
Analysis of Fig.\ 11 of Gnedin \& Hui \shortcite{GH97} shows that the HPM
introduces about 3\% error to both the slope and the amplitude of the power
law. Analysis of Fig.\ \ref{figVP} also shows that a 3\% error in the slope
and a 2\% error in the amplitude is introduced by using the DPA instead
of Voigt profile fitting for column densities between 
$10^{13}\dim{cm}^{-2}$ and $10^{14}\dim{cm}^{-2}$. Finally, due to the finite
size of a computational box, only a subset of all possible initial conditions
can be simulated (only one ``realization''), and this statistical
undersampling also introduces an error. In order to estimate the error
due to the realization dependence, I have performed three different
realizations of the same cosmological model
with the $256^3$ mesh and two realizations with the $512^3$ mesh
(models SCDM.2A, SCDM.2D, and SCDM.2E).

\def\capRD{
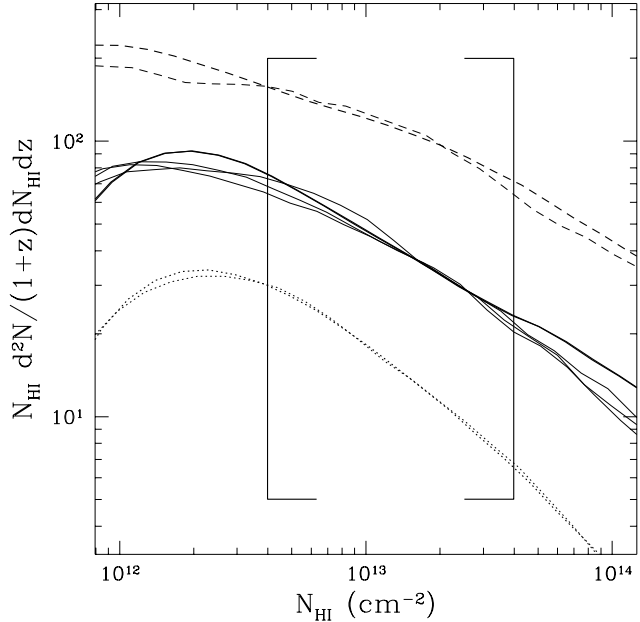
Column density distributions for three variants of the SCDM model: $\sigma_8
=0.5$ (dotted lines), $\sigma_8=0.7$ (solid lines), and $\sigma_8=1.2$
(dashed lines). Thin lines show $256^3$ simulations and bold lines show
$512^3$ simulations. For $\sigma_8=0.7$ model 
three different random realization
with $256^3$ resolution and two different random realizations with 
$512^3$ resolution
are shown.
The square bracket shows the region where
numerical convergence with respect to the box size is achieved.
Lines of different type are offset vertically for the sake of clarity.
}
\begin{figure}
\par\centerline{%
\epsfxsize=1.0\columnwidth\epsfbox{\figname{figRD.ps}}}%
\caption{\label{figRD}\capRD}
\end{figure}
Figure \ref{figRD} shows column density distributions for all realizations,
normalized as explained in \S \ref{normalexp}.
The difference in slopes and in power law
amplitudes for three $256^3$ realizations reaches 10\% and 18\% respectively,
while $512^3$ simulations have an error a factor of three less in both
the slope and the amplitude.
Thus, performing $512^3$ simulations for all 25 model would substantially
reduce the theoretical error, but, as I mention above, since this is beyond
the practically possible at the moment,
the total error from Table \ref{taberr} should
be properly included in all comparisons between the simulations and 
the observations.

While Table \ref{taberr} gives a magnitude of the error, it does not
determine the statistical property of this error. In other words, it
does not specify what the probability is that a given model would produce
results accurate to within the error from Table \ref{taberr}. While there
is no simple way to find out the probability distribution of the error,
the error in Table \ref{taberr} should be considered as about 95\%
confidence interval from the way it was computed. This assertion is also
supported by comparison between the observations and the simulations
in the following section.

\subsection{Observational data}
\label{normalexp}

I now turn to directly comparing the results of the simulations with the
observational
data. Recent Keck observations of the Lyman-alpha forest 
\cite{HKCS95,LSWT96,KT97,KHCS97} give the column density distribution
of the Lyman-alpha forest
at four different redshifts: $z=2.31$, $z=2.85$, $z=3.35$, and $z=3.70$.
The data from Kirkman \& Tytler \shortcite{KT97} have a 
median redshift of $z=2.70$ and about 10\%
higher amplitude than the data from Hu et al.\ \shortcite{HKCS95}.
Some part of this difference is due to the slightly different redshift
of Kirkman \& Tytler observations, and some other part (not quite certain
at the moment) is due to the Cosmic Variance. But since
the current error of my simulations is about 20\% in the amplitude, 
both data samples are consistent with each other for my purpose.

\begin{table}
\caption{Column density limits}
\label{tablim}
\medskip
\begin{tabular}{cccc}
\noalign{\hrule\smallskip}
Redshift & 
${\rm lg}(N_{\HI,{\rm MIN}})$ & 
${\rm lg}(N_{\HI,{\rm MAX}})$ \\ 
\noalign{\smallskip\hrule\smallskip}
2.31	& 12.6 & 13.2 \\
2.85	& 12.6 & 13.6 \\
3.35	& 13.0 & 14.0 \\
3.70	& 13.4 & 14.0 \\
\noalign{\smallskip\hrule}
\end{tabular}
\end{table}
As Fig.\ \ref{figRD} also demonstrates, only within a limited range of
column densities there is a consistency between $256^3$ and $512^3$
simulations, i.e.\ only with this final range a numerical convergence is
achieved. Thus, only this range of column densities can be used in
comparing observations and simulations. In addition, this range is
also a function of redshift, and a figure similar to Fig.\ \ref{figRD}
at each value of redshift is used to determine the range of column
densities at each redshift within which a numerical convergence is
achieved. Whenever the range of convergence exceeds the range
of applicability of the HPM-DPA approximation, a smaller of two ranges
is used (thus maintaining both the accuracy of the HPM-DPA approximation
and the numerical convergence of simulations). As can be seen from
Fig.\ \ref{figRD} the range of convergence is a weak function of the
amplitude of the density fluctuations, and thus the same range of column
densities can be
used for all models.

The specific limits are summarized in Table \ref{tablim}.
I only point out here that $512^3$ simulations have not been continued
beyond $z=2.85$, and the fitting interval at $z=2.31$ is determined
by extrapolation of fitting intervals at higher redshifts. One thus
has to keep in mind that numerical convergence has not been firmly
established for simulation results at $z=2.31$. Therefore, I will
only use simulation results at $z=2.85$, $z=3.35$, and $z=3.70$ for
quantitative comparison with observations, and will use model
predictions at $z=2.31$ for illustrative purposes only.

For a given range of column densities, I can fit a power law
(simultaneously in the slope and amplitude)
to both the results of simulations and the observational data. I then
fix the amplitude of the column density distribution from the
simulation by requiring that it coincides with the amplitude
fitted to the observational data at the specific value of the
column density $N_{\HI,{\rm FIT}}=3\times10^{13}\dim{cm}^{-2}$ (for
all four values of redshift). 
Thus, for a given simulation
this procedure produces two numbers: the slope of the column
density distribution $\beta$ and the amplitude of the column density
distribution, which can be expressed as $J_{21}$.
It is the time
evolution and the dependence on cosmological parameter
of those two quantities
that I concentrate on in this paper.

\section{Results}

\subsection{The slope of the column density distribution}

I now focus on investigating the dependence of the slope of the
column density distribution $\beta$ on cosmological parameters.

\def\capBS{
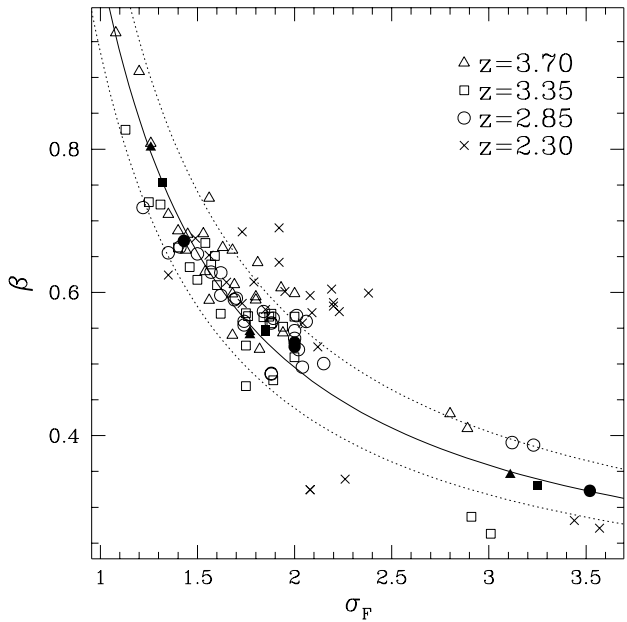
The slope of the column density distribution as a function of the
rms linear density fluctuation at the filtering scale 
$\sigma_{F}$ for all
models at four different redshifts: 
$z=2.31$ (crosses),
$z=2.85$ (circles),
$z=3.35$ (squares),
$z=3.70$ (triangles). Open symbols mark 25 $256^3$ simulations, and
filled symbols show 4 $512^3$ simulations. The solid line shows the 
fit \protect{\ref{betabar}}, and dotted lines show $\pm13\%$ deviation
from the fit.
}
\begin{figure}
\par\centerline{%
\epsfxsize=1.0\columnwidth\epsfbox{\figname{figBS.ps}}}%
\caption{\label{figBS}\capBS}
\end{figure}
Figure \ref{figBS} presents the major result of this paper, the
dependence of the slope of the column density distribution on the
rms linear density fluctuation at the filtering scale. Various open
symbols show
computed power law slopes for all 25 $256^3$ models at four 
different redshifts, and respective solid symbols show four $512^3$
models at three highest redshifts. The solid
line shows the fit in the form
\begin{equation}
	\bar\beta(\sigma_F) = {0.76\over {\rm ln}(1+\sigma_F^2)^{0.9}},
	\label{betabar}
\end{equation}
and dotted lines mark $\pm13\%$ interval around the fit
(which is consistent with being a 95\% confidence interval). 
I conclude that
within the errors of the approximation, the slope of the column density
distribution is a function of {\it the rms linear density fluctuation at
the filtering scale\/} alone. This conclusion quantitatively confirms
the qualitative conclusion of Hui et al.\ \shortcite{HGZ97}, who found 
that more nonlinear models
have shallower column density distributions. 

The scatter is much smaller for the $512^3$ models, as should be
expected from Table \ref{taberr}. However, since four $512^3$ do
not cover the full space of cosmological parameters, they cannot be
used to investigate whether the slope of the column density distribution
depends on other parameters in addition to $\sigma_F$. I however
use them to accurately determine the best fit. Fig.\ \ref{figBS}
thus demonstrates that the slope of the column density distribution of
the Lyman-alpha forest $\beta$ at $N_{\HI,{\rm FIT}}=
3\times10^{13}\dim{cm}^{-2}$ is
the function of the density fluctuation on the filtering scale alone
to within 13\%. It is very likely, judging from how well points
from $512^3$ simulation fit the solid line, that $\beta$ is a function
of $\sigma_F$ to much better accuracy (at least 5\%), but to
verify this requires running several more $512^3$ simulations. This
work in now in progress.

The scatter around the fit (\ref{betabar}) may be caused by two different
effects: by the error in the HPM-DPA calculation, and by dependence
of the slope $\beta$ on other cosmological parameters, like the local
slope of the power spectrum, the equation of state etc. Because the scatter
is consistent with the error alone, current work is not able to address
the second source of scatter, i.e.\ it is possible that the slope of the
column density distribution does depend on other parameters of the model,
but this dependence is weaker than 13\% and therefore cannot be uncovered
here. I would like to emphasize that this conclusion does not necessarily
contradict the conclusion of Hui et al.\ \shortcite{HGZ97}, who found 
a strong dependence
of the slope of the column density distribution on both the local
slope of the density power spectrum and on the equation of state slope
$\gamma$. The conclusion of Hui et al.\ \shortcite{HGZ97} was based on 
using a quantity
called $\sigma_0$ as a measure of the nonlinearity of the model. This
quantity is ill-defined, because it measures the amplitude of linear density
fluctuation on a scale which is model-dependent and lacks any physical
explanation, while $\sigma_F$ represents to a high degree of accuracy
(as have been shown in Gnedin \& Hui 1997) the real total (i.e.\
summed over all scales) rms linear gas density fluctuation. I also note here 
that since
the Zel'dovich approximation used by Hui et al.\ \shortcite{HGZ97} is 
uncontrolled approximation, its accuracy is not known.

I also note here that the scatter around the fit gets larger for higher
values of $\sigma_F$, as could be expected, as higher $\sigma_F$ corresponds
to more nonlinear models, for which the cosmic variance is more significant.
The crosses, which correspond to $z=2.31$, have a larger scatter, and
also somewhat biased toward higher values of $\beta$. This is due to
the fact that the range of column densities within which numerical 
convergence is achieved is not established at $z=2.31$, as have been
explained in the previous section. I show the simulation results
at $z=2.31$ here for illustration purposes only; they are not used to 
determine the fit (\ref{betabar}).

Fig.\ \ref{figBS} presents the slope $\beta$ as a function of 
$\sigma_{F,{\rm BOX}}$, because it is $\sigma_{F,{\rm BOX}}$, rather
than $\sigma_{F}$, which is actually computed in a simulation.
But now, given the dependence of the slope $\beta$ on $\sigma_F$, it is
possible to correct for this bias, and, given the slope of the column density
distribution derived from a simulation, $\beta_{\rm SIM}$, to compute the 
slope $\beta_{\rm COR}$ which would be
derived in an idealistic simulation of an infinite size:
\begin{equation}
	\beta_{\rm COR} = \beta_{\rm SIM} + \bar\beta(\sigma_F) -
	\bar\beta(\sigma_{F,{\rm BOX}}),
	\label{betacor}
\end{equation}
and correct the column density distribution itself accordingly:
\[
        \left.{N_{\HI}d^2{\cal N}\over (1+z)dz\,dN_{\HI}}\right|_{\rm COR}
        =
\]
\begin{equation}
        \ \ \ \ 
        \left.{N_{\HI}d^2{\cal N}\over (1+z)dz\,dN_{\HI}}\right|_{\rm SIM}
        \left(N_{\HI}\over N_{\HI,{\rm FIT}}\right)^
        {\bar\beta(\sigma_F)-\bar\beta(\sigma_{F,{\rm BOX}})}.
        \label{cddcor}
\end{equation}
where $N_{\HI,{\rm FIT}}=3\times10^{13}\dim{cm}^{-2}$.

After having corrected for the bias due to the finite simulation size,
I can directly compare cosmological models to the observations. Let me first
concentrate on $z=2.85$, since for this value of redshift the HPM-DPA
calculation is the most accurate. The observed slope of the column density
distribution is $\beta=0.5$ \cite{HKCS95,KT97}. However, the error of this
number is not quite certain. Therefore, I will say that the model fits the
observations if its slope $\bar\beta$ is between 0.4 and 0.6
(20\% total [15\% observational plus 13\% modeling] error); 
if the slope of the column density distribution
for a model falls between 0.35 and 0.40 or between 0.60 and 0.65
(30\% total error), I will say that the model
marginally fits the data; finally, if the slope $\beta$ for a given model
is above 0.65 or below 0.35, I will assume that the model fails.

\begin{table*}
\begin{minipage}{\hsize}
\caption{Models vs observations}
\label{tabfit}
\begin{tabular}{llllllllllrc}
\noalign{\hrule\smallskip}
Model & $\Omega_0$ & $\Omega_b$ & $\Omega_\Lambda$ & $\Omega_{\nu}$ & $h$ 
& $n$ & $\sigma_8$ & $\beta_{\rm SIM}$ & $\beta_{\rm COR}$& $\bar\beta$ & fit? \\
\noalign{\smallskip\hrule\smallskip}
SCDM.5A$^{\rm a}$ 
        & 1    & 0.05  & 0    & 0    & 0.5  & 1    &  1.2  & 0.33 & 0.32 & 0.30 & $  -  $ \\
TCDM.2A & 1    & 0.05  & 0    & 0    & 0.5  & 0.9  &  0.83 & 0.57 & 0.51 & 0.48 & $  +  $ \\
LCDM.1A & 0.35 & 0.05  & 0.65 & 0    & 0.7  & 1    &  0.94 & 0.59 & 0.51 & 0.51 & $  +  $ \\
LCDM.1C & 0.35 & 0.05  & 0.65 & 0    & 0.7  & 1    &  0.94 & 0.56 & 0.52 & 0.44 & $  +  $ \\
\noalign{\smallskip}
LCDM.2A & 0.35 & 0.05  & 0.65 & 0    & 0.7  & 0.96 &  0.80 & 0.63 & 0.56 & 0.55 & $  +  $ \\
LCDM.3A & 0.35 & 0.03  & 0.65 & 0    & 0.7  & 1    &  1.04 & 0.57 & 0.51 & 0.44 & $  +  $ \\
LCDM.4A & 0.35 & 0.03  & 0.65 & 0    & 0.7  & 0.95 &  0.85 & 0.65 & 0.56 & 0.58 & $  +  $ \\
LCDM.5A & 0.4  & 0.036 & 0.6  & 0    & 0.65 & 1    &  1.02 & 0.54 & 0.48 & 0.44 & $  +  $ \\
CHDM.1A & 1    & 0.05  & 0    & 0.1  & 0.5  & 1    &  0.92 & 0.52 & 0.47 & 0.44 & $  +  $ \\
\noalign{\smallskip}
CHDM.2A & 1    & 0.05  & 0    & 0.15 & 0.5  & 1    &  0.86 & 0.63 & 0.54 & 0.56 & $  +  $ \\
CHDM.3A & 1    & 0.05  & 0    & 0.2  & 0.5  & 1    &  0.82 & 0.72 & 0.57 & 0.73 & $  -  $ \\
CHDM.4A & 1    & 0.05  & 0    & 0.1  & 0.65 & 1    &  1.22 & 0.39 & 0.37 & 0.31 & $  -  $ \\
CHDM.5A & 1    & 0.05  & 0    & 0.2  & 0.65 & 1    &  1.12 & 0.56 & 0.49 & 0.46 & $  +  $ \\
CHDM.6A & 1    & 0.07  & 0    & 0.1  & 0.6  & 0.9  &  0.73 & 0.66 & 0.56 & 0.63 & $ \pm $ \\
\noalign{\smallskip\hrule\medskip}
\end{tabular}
\end{minipage}
$^{\rm a}$ A $512^3$ simulation is used for this model.\hfill\break
\end{table*}
Table \ref{tabfit} presents the biased $\beta_{\rm SIM}$ and corrected 
$\beta_{\rm COR}$ values for the slope of the column density distribution
for all {\it COBE\/} normalized models from Table \ref{tabmod}, as well
as the value of $\bar\beta$ computed from equation (\ref{betabar}).
The cosmological parameters for the models are included for reference.
Finally, the last column indicates whether the model
fits the data ($+$), marginally fits ($\pm$), or fails completely ($-$).

Finally, using equation (\ref{betabar}), it is possible to place a constraint
on the value of the rms linear density fluctuation $\sigma_F$. The 
observational constraint $0.4<\beta<0.6$ can be expressed in terms of $\sigma_F$
as
\begin{equation}
	1.6 < \sigma_F < 2.6.
	\label{sigfconst}
\end{equation}
This should be considered as roughly 2 $\sigma$ (95\% confidence level) interval.
With the 3 $\sigma$ error, this interval becomes
\[
	1.5 < \sigma_F < 3.1.
\]
The median value, $\beta=0.5$ corresponds to $\sigma_F=2.0$.

Finally, as I have mentioned in the previous section, the filtering scale
$1/k_F$ depends on reionization history and is a nontrivial function of
time. For a purpose of a quick estimate, it would be more convenient to use
a fixed scale like $1/k_{34}$ (eq. [\ref{ktfdef}]). The following approximate
expression accurate to about 5\% can then be used to compute $\sigma_F$
at $z=2.85$ from $\sigma_{34}$:
\[
	\sigma_F(z=2.85) = \sigma_{34}
	\left[
	1.06 - 0.9(\gamma-1.5) - 0.15(T_4-1)\right].
\]
This expression depends on the parameters of the equation of state since
$\sigma_F$ depends on the equation of state while $\sigma_{34}$ does not.
I point out here that the above expression is derived by analyzing 25
models considered in this paper, and may not be valid for a model that falls
outside the range of cosmological models considered here, or for a model
whose equation of state evolves significantly differently from the analytical
expressions of Hui \& Gnedin \shortcite{HG97}.

\subsection{The amplitude of the column density distribution}

While the slope of the column density distribution can serve as a basis
for discriminating against some of the cosmological models, the amplitude
of the column density distribution depends on the assumed value for
the ionizing intensity $J_{21}$, which is currently highly uncertain.
The best available measurements are based on the proximity effect, and
give values in the range $J_{21}=0.2-1.5$ at $z=2-4$
\cite{Gea96,LSWT96,CEC97}. I therefore consider $J_{21}$ as a free parameter,
and use the observational data to fix the value for $J_{21}$ that best
fits the data.

Let me first consider $z=2.85$. The normalization procedure described in the
previous subsection defines a value of $J_{21}$ for each cosmological model.
After computing the respective $J_{21}$ values for all 25 models, I obtain
the following {\it approximate\/} expression for $J_{21}$ as a function of
cosmological parameters:
\begin{equation}
	J_{21}(z=2.85) = 
	10^3 {\Omega_b^2 h^3\over \Omega_0^{1/2}T_4^{0.7} \sigma_F},
	\label{ampmed}
\end{equation}
which is accurate to about 40\%. The accuracy estimate results form the
assumed accuracy of 20\% for the amplitude of the column density
distribution, which in turn translates into a 40\% accuracy in 
$J_{21}\propto N_{\HI}$ for the power-law column density distribution
with the index $\beta\approx0.5$. 

\def\capAA{
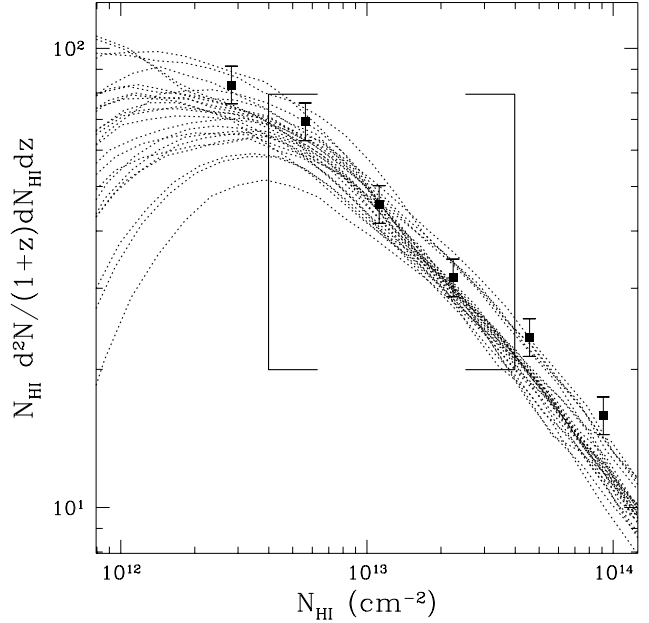
Comparison between the observational data at $z=2.85$ and all 25 cosmological 
models, reduced to have the slope of the column density distribution of
0.5 (eq.\ [\protect{\ref{cddred}}])
and normalized according to equation \protect{\ref{ampmed}}
(dotted lines).
}
\begin{figure}
\par\centerline{%
\epsfxsize=1.0\columnwidth\epsfbox{\figname{figAA.ps}}}%
\caption{\label{figAA}\capAA}
\end{figure}
In order to demonstrate graphically the level of accuracy of this 
approximation, I plot all 25 models with the value of $J_{21}$ defined
according to equation (\ref{ampmed}), ``reduced'' to have 
the same slope of the column density distribution, 
\[
	\left.{N_{\HI}d^2{\cal N}\over (1+z)dz\,dN_{\HI}}\right|_{\rm RED}
	\equiv 
\]
\begin{equation}
	\ \ \ \
	\left.{N_{\HI}d^2{\cal N}\over (1+z)dz\,dN_{\HI}}\right|_{\rm SIM}
	\left(N_{\HI}\over N_{\HI,{\rm FIT}}\right)^
	{0.5-\bar\beta(\sigma_{F,{\rm BOX}})}.
	\label{cddred}
\end{equation}
In other words, if both equation (\ref{betabar}) and equation (\ref{ampmed})
were exact, the reduced column density distribution would be the same
for all cosmological models. Thus, the difference between different lines
in Fig.\ \ref{figAA} demonstrates the error in both  
equation (\ref{betabar}) and equation (\ref{ampmed}).

One can note that while the majority of models lie together in a strip
coinciding with the data points, there are six models
(SCDM.2D, SCDM.2G, SCDM.2H, SCDM.2L, LCDM.1C and LCDM.2A) that lie above the rest. This
difference correlates with neither of the cosmological parameters
defining the model. More than that, since one of the outliers 
(SCDM.2D) is just a 
different random realization of the SCDM.2A model, the difference in
the amplitude is real, and is due to the error induced by the HPM-DPA 
approximation,
most likely the finite size of a simulation box. One therefore may hope that
the dispersion in Fig.\ \ref{figAA} can be reduced by using larger simulation
sizes (work currently in progress).

Finally, I would like to point out here that $\Omega_b$ and $h$
dependence in equation (\ref{ampmed}) agrees with the one obtained
by Croft et al.\ \shortcite{CWKH96} when the dependence of $T_0$ on
$\Omega_b$ (for a given cosmological model) is taken into account
\cite{HG97}. However,
the advantage of equation (\ref{ampmed}) is that it is (approximately)
valid for {\it any\/} cosmological model, i.e.\ there exist no additional
dependences on physical parameters that are not incorporated into
equation (\ref{ampmed}), except at a level comparable to or below
the error of the HPM-DPA calculation.

I note here, that since both $T_0$ and $\sigma_F$ are non-trivial functions of 
time, and it is unlikely that $z=2.85$ represents a special moment in the
history of the universe, equation (\ref{ampmed}) cannot hold at some other
redshift with the coefficient $10^3$ being replaced by some other appropriate
number. In other words, while it is plausible that the expression (\ref{betabar})
for the slope of the column density distribution is accurate to high precision,
equation (\ref{ampmed}) is {\it necessary approximate\/}, and 
there {\it has\/} to be some other dependence in equation (\ref{ampmed}) on 
cosmological parameters at the 40\% or below level.

\subsection{The evolution of the column density distribution}

Using the observational data at four different redshifts, it is possible
to constraint the evolution of the column density distribution of the
Lyman-alpha forest from $z\sim4$ to $z\sim2$.

\def\capEU{
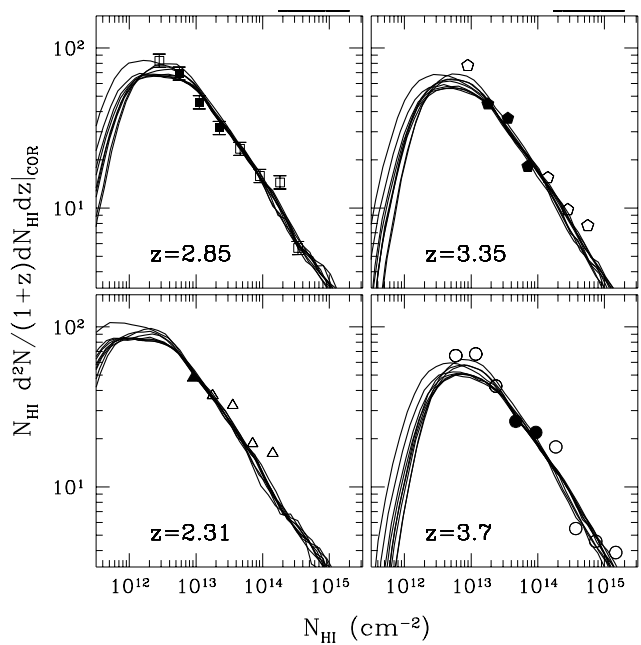
Comparison between the models that fit the data at $z=2.85$ and observations
at four different redshifts. The same value of $J_{21}$ is adopted for each 
cosmological model at all four redshifts. Symbols show the observational
data with solid symbols lying in the column density range used to fit
the amplitude of the column density distribution.
}
\begin{figure}
\par\centerline{%
\epsfxsize=1.0\columnwidth\epsfbox{\figname{figEU.ps}}}%
\caption{\label{figEU}\capEU}
\end{figure}
I show in Figure \ref{figEU} the column density distributions for the
seven models from Table \ref{tabfit} that fit the data at all four redshifts.
For every model the value of $J_{21}$ is the same at each redshift
(but different for different models). Since
all models are consistent with the data at all four redshifts, the observations
are consistent with the value of $J_{21}$ being independent of redshift
in the redshift interval $z\sim2-4$.
This normalization is also 
fully consistent with the slightly higher abundance of the Lyman-alpha
forest absorbers at $z\approx3$ found by Kirkman \& Tytler \shortcite{KT97}
within the errors of the analysis presented in this paper.

\section{Discussion}

I have demonstrated that the slope of the column density distribution
of the Lyman-alpha forest
is a function of the rms linear density fluctuation of the gas distribution
(which is smooth on scales below the characteristic filtering scale), and
is independent of other cosmological parameters within the accuracy of
about 13\% of the approximate method utilized (Hydro-PM approximation 
combined with the Density-Peak Ansatz). The amplitude of the column
density distribution, expressed in terms of the ionizing intensity
$J_{21}$, is proportional to a combination of cosmological parameters,
\[
	J_{21}\propto
	{\Omega_b^2 h^3\over \Omega_0^{1/2}T_4^{0.7} \sigma_F}, 
\]
and is independent of other cosmological or physical parameters within
about 40\% accuracy, at a given value of redshift.

The observational data and the modeling technology are currently consistent
with the value of $J_{21}$ being constant from $z\sim4$ until 
$z\sim2$.

In order for a cosmological model to fit the observed slope of the column
density distribution of the Lyman-alpha forest at $z\approx3$, the rms linear
density fluctuation at the filtering scale 
$k_F\sim k_{34}$ (see eq.\ [\ref{ktfdef}]) in the model should be between
1.6 and 2.6 (or between 1.5 and 3.1 with a conservative estimate).

Why is the rms linear density fluctuation the major factor determining the
slope of the column density distribution, when $\sigma_F\sim2$ means
an already nonlinear distribution? The answer is that $\sigma_F$ serves
as {\it a\/} measure of the nonlinearity of the model; one could have
adopted another measure, say, the total (nonlinear) rms density fluctuation
in the model. However, since the total rms density fluctuation is a function
of the rms linear density fluctuation \cite{Kea94} with, perhaps, 
a slight dependence on the local value of the slope of the density 
power spectrum, which is within the uncertainty of the simulations presented
here, any measure of the nonlinearity of the model can be expressed as
a function of $\sigma_F$.

Finally, a few comments can be made about fitting a cosmological model
to the observational data. If one ignores semi-philosophical questions
about the Hubble constant, the age of the universe, and the value of
$\Omega_0$, and restricts oneself to requiring that a given cosmological model 
satisfy
three measurements of the underlying density power spectrum: the {\it COBE\/}
normalization, the cluster abundance, and the slope of the column density
distribution of the Lyman-alpha forest at $z\approx3$, once can then ask
a question how well the cosmological models listed in Table \ref{tabmod}
would perform on those three tests. 
Let me concentrate only on the models that pass the first test,
i.e.\ only on the {\it COBE\/} normalized models listed in Table \ref{tabfit}.
Using the fit $\sigma_{8,{\rm CLUSTER}}$
for the top-hat linear density fluctuation at the $8h^{-1}\dim{Mpc}$ scale
that satisfies the cluster constraint from Eke, Cole, \& Frenk 
\shortcite{ECF96}, 
\[
	\sigma_{8,{\rm CLUSTER}} = (0.50\pm0.04)\Omega_0^{-0.53+0.13\Omega_0},
\]
I compute the error in the amplitude of the linear density
fluctuation at the cluster scale,
\[
	{\sigma_8\over\sigma_{8,{\rm CLUSTER}}}-1.
\]
The error in the slope of the column density distribution of the Lyman-alpha
forest (assuming the observations give $\beta=0.5$) is simply
\[
	{\beta\over0.5}-1.
\]
\def\mpass{\ \ \ \ \surd}
\def\pass{$\mpass$}
\def\fail{$\phantom{\mpass}$}
\begin{table*}
\begin{minipage}{\hsize}
\caption{{\it COBE\/} normalised models}
\label{tabcon}
\begin{tabular}{llllllllrr}
\noalign{\hrule\smallskip}
Model & $\Omega_0$ & $\Omega_b$ & $\Omega_\Lambda$ & $\Omega_{\nu}$ & $h$ 
& $n$ & $\sigma_8$ & ${\displaystyle\sigma_8\over\displaystyle\sigma_{8,{\rm CLUSTER}}}-1$ & ${\displaystyle\beta\over\displaystyle0.5}-1$\ \ \  \\
\noalign{\smallskip\hrule\smallskip}
SCDM.5A & 1    & 0.05  & 0    & 0    & 0.5  & 1    &  1.20 & +140\%\fail &  -40\%\fail \\
TCDM.2A & 1    & 0.05  & 0    & 0    & 0.5  & 0.9  &  0.83 &  +66\%\fail &   -4\%\pass \\
LCDM.1A & 0.35 & 0.05  & 0.65 & 0    & 0.7  & 1    &  0.94 &  +13\%\pass &   +2\%\pass \\
LCDM.1C & 0.35 & 0.05  & 0.65 & 0    & 0.7  & 1    &  0.94 &  +13\%\pass &  -12\%\pass \\
\noalign{\smallskip}
LCDM.2A & 0.35 & 0.05  & 0.65 & 0    & 0.7  & 0.96 &  0.80 &   -4\%\pass &  +10\%\pass \\
LCDM.3A & 0.35 & 0.03  & 0.65 & 0    & 0.7  & 1    &  1.04 &  +25\%\fail &  +12\%\pass \\
LCDM.4A & 0.35 & 0.03  & 0.65 & 0    & 0.7  & 0.95 &  0.85 &   +2\%\pass &  +16\%\pass \\
LCDM.5A & 0.4  & 0.036 & 0.6  & 0    & 0.65 & 1    &  1.02 &  +32\%\fail &  -12\%\pass \\
CHDM.1A & 1    & 0.05  & 0    & 0.1  & 0.5  & 1    &  0.92 &  +84\%\fail &  -12\%\pass \\
\noalign{\smallskip}
CHDM.2A & 1    & 0.05  & 0    & 0.15 & 0.5  & 1    &  0.86 &  +72\%\fail &  +46\%\fail \\
CHDM.3A & 1    & 0.05  & 0    & 0.2  & 0.5  & 1    &  0.82 &  +64\%\fail &  +64\%\fail \\
CHDM.4A & 1    & 0.05  & 0    & 0.1  & 0.65 & 1    &  1.22 & +144\%\fail &  -38\%\fail \\ 
CHDM.5A & 1    & 0.05  & 0    & 0.2  & 0.65 & 1    &  1.12 & +124\%\fail &   -8\%\pass \\
CHDM.6A & 1    & 0.07  & 0    & 0.1  & 0.6  & 0.9  &  0.73 &  +46\%\fail &  +26\%\fail \\
\noalign{\smallskip\hrule\smallskip}
\end{tabular}
\end{minipage}
\end{table*}
Table \ref{tabcon} lists all fourteen models together with the respective
 errors
in the density fluctuation amplitude at the cluster scale and in the slope
of the column density distribution of the Lyman-alpha forest
(the check mark after the error shows
whether the model passes the test at a 95\% confidence level or not).
One can see that many models do well. Several of LCDM models pass
all three tests. The fact that none of the CHDM models presented
pass all the tests does not imply that this model can be ruled out:
it seems plausible that a set of parameters can be found that will
satisfy all three tests. The set of CHDM models used in this paper
has been chosen ``randomly'', without prior attempts to make these
models fit the cluster abundance and the slope of the column density
distribution tests. The fact that several of ``randomly'' chosen
LCDM models perform well implies that there is more parameter space
available for the LCDM models than for CHDM models, but it will be
premature to claim that CHDM model fail altogether.

It however seems possible to make a claim that the density parameter
in massive neutrinos, $\Omega_\nu$, has to be lower than or about $0.15$.
The two models, CHDM.3A and CHDM.5A, represent two extreme cases of
the $\Omega_\nu=0.2$ models, and they clearly demonstrate that
models with $\Omega_\nu=0.2$ have too little power on scales of about
$100\dim{kpc}$ as compared with $10\dim{Mpc}$ scales.

I am very grateful to Romeel Dave for letting me use his AUTOVP automated
Voigt profile fitting software. This work was supported by the 
UC Berkeley grant 1-443839-07427.
Simulations were performed on the NCSA Power Challenge Array under 
the grant AST-960015N and on the NCSA Origin2000 mini 
super-computer under the grant AST-970006N.

\end{document}